\documentclass[a4paper,11pt]{article}
\usepackage{pos}

\newcommand{\be}{\begin{equation}}
\newcommand{\ee}{\end{equation}}
\newcommand{\ba}{\begin{eqnarray}}
\newcommand{\ea}{\end{eqnarray}}
\newcommand{\nn}{\nonumber}

\title{Possible $K \bar{K}^*$ and $D \bar{D}^*$
 resonances by solving Schr\"odinger equation}

\author*[a]{Bao-Xi Sun}
\author[a]{Qin-Qin Cao}
\author[b]{Ying-Tai Sun}

\affiliation[a]{School of Physics and Optoelectronic Engineering, Beijing University of Technology, \\
Beijing 100124, China}

\affiliation[b]{School of Mechanical and Materials Engineering, North China University of Technology,  \\ Beijing 100144, China}

\emailAdd{sunbx@bjut.edu.cn}

\abstract{The one-pion exchange interaction between the kaon and the vector antikaon is investigated by solving the Schr\"odinger equation in the S-wave approximation. In addition to the particle $f_1(1285)$, another bound state of $K \bar{K}^*$ is obtained, which is approximately 9 MeV below the threshold of $K \bar{K}^*$ and labeled $f_1(1378)$ for convenience in this manuscript. Under the outgoing wave condition, two resonance states of $K \bar{K}^*$  are produced with different coupling constants fixed with the binding energies of $f_1(1285)$ and $f_1(1378)$, respectively. Both of the resonance states are located in the vicinity of 1400 MeV, and thus it is reasonable to assume that these two resonance states correspond to the $f_1(1420)$ particle in the review of the Particle Data Group simultaneously.
This method is extended to study the $D \bar{D}^*$ system analogously. When the particle $\chi_{c1}(3872)$ is treated as a bound state of $D \bar{D}^*$, the particles $T_{c\bar{c}1}(3900)$, $T_{c \bar{c}}(4020)$ and $X(3940)$ can be obtained as solutions of the Schr\"odinger equation under the outgoing wave condition, which implies the spin and parity of these particles are all $J^P=1^+$.}

\FullConference{The 21st International Conference on Hadron Spectroscopy and Structure (HADRON2025)\\
 27 - 31 March, 2025\\
Osaka University, Japan\\}

\begin{document}
\maketitle

\section{Introduction}
\label{sect:Introduction}

The $K \bar{K}^*$ system is investigated by solving the Schr\"odinger equation under the outgoing wave condition, and a complex solution above the threshold of $K \bar{K}^*$ is obtained, which is assumed to be the $f_1(1420)$ particle in the review of the Particle Data Group(PDG)\cite{PDG2024}. When the Schr\"odinger equation is solved, an one-pion-exchange potential of the kaon and the vector antikaon is adopted, which is different from the kernel used in the unitary coupled channel approximation, where a vector meson exchange is dominant according to the SU(3) hidden guage 
symmetry\cite{Roca2005,sun1420}. Reasonably, this method is extended to study the $D \bar{D}^*$ case\cite{sunbx2024}.

In the next section, descriptions of the Schr\"odinger equation with an one-pion-exchange potential will be reviewed according to Ref.~\cite{sunbx2024}. In Sections~\ref{sect:KKstar} and~\ref{sect:DDstar} , the calculation results for the bound and resonance states of $K \bar{K}^*$ and $D \bar{D}^*$ will be evaluated, respectively. Finally, Section~\ref{sect:summary} is devoted to a brief summary. 

\section{Framework}
\label{sect:framework}

If the interaction of the kaon and the vector antikaon is realized by exchanging a pion, the potential of them can be indicated as  
\be
\label{eq:202307071816}
V(r)=-g^2\frac{e^{-mr}}{d},
\ee
where $m$ is the mass of the pion, $g$ is the coupling constant, and the distance $r$ in the denominator has been replaced approximately with the force range $d=1/m$.
It is apparent that the potential in Eq.~(\ref{eq:202307071816}) is reasonable in the force range, and it is equal to the Yukawa-type potential asymptotically at infinity. Under this approximation, the Schr\"odinger equation can be solved analytically.

If the radial wave function takes the form of  $R(r)=\frac{u(r)}{r}$, the radial Schr\"odinger equation in the S-wave approximation can be written as
\be
\label{eq:202307081218}
-\frac{\hbar^2}{2\mu} \frac{d^2 u(r)}{dr^2}+V(r)u(r)=Eu(r),
\ee
where $\mu$ is the reduced mass of the two-body system.

According to the substitution of variables 
\be 
u (r) = J (x), \nn
\ee
and
\be
r \rightarrow x=\alpha e^{-\beta r},~~~~0 \le x \le \alpha, \nn
\ee
with
\be
\alpha=2g \sqrt{2 \mu d},~~~~\beta=\frac{1}{2d}, \nn
\ee
and
\be
\label{eq:rhoenergy}
\rho^2=-8d^2 \mu E,~~~~E \le 0,
\ee
the radial Schr\"odinger equation in Eq.~(\ref{eq:202307081218}) becomes the $\rho$th order Bessel equation, 
\be
\label{eq:202307081903}
\frac{d^2 J(x)}{d x^2}+\frac{1}{x} \frac{d J(x)}{d x} +\left[1-\frac{\rho^2}{x^2}\right] J(x)=0, \nn
\ee
and its solution is the $\rho$th order Bessel function $J_\rho(x)$.

For the bound state, when $r \rightarrow +\infty$, the radial wave function $R(r) \rightarrow 0$, and thus $u(r)=J_\rho(\alpha e^{-\beta r})=J_\rho(0)$ with $\rho \ge 0$. On the other hand, when $r \rightarrow 0$, $u(r) \rightarrow 0$, and it means
\be
\label{eq:Besselzeropoint}
J_\rho(\alpha)=0.
\ee
Therefore, if only one bound state of $K\bar{K}^*$ has been detected, the order of the Bessel function $\rho$ in Eq.~(\ref{eq:Besselzeropoint}) can be determined with the binding energy $E$, and then the coupling constant $g$ in the Yukawa-type potential is obtained with the first nonzero zero point of the Bessel function $J_\rho(\alpha)$, which takes a form of
\be
\label{eq:couplingzeropoint}
g^2=\frac{\alpha^2}{8 \mu  d}.
\ee

The second kind of Hankel function $H_\rho^{(2)}(x)$ is also an independent solution of the Bessel equation. Under the outgoing wave condition
\be
\label{eq:outgoing}
H_\rho^{(2)}(\alpha)=0,
\ee
the $K\bar{K}^*$ resonance state can be obtained when the value
$\alpha$ is fixed with Eq.~(\ref{eq:Besselzeropoint}).
Apparently, the energy of the resonance state is related to the order of the second kind of Hankel function and takes a complex value
of $E=M-i\frac{\Gamma}{2}$, where the real part represents the mass of the resonance state, and the imaginary part is one half of the decay width, $\Gamma=-2iImE$, as discussed in Ref.~\cite{Moiseyev}.

\section{$K \bar{K}^*$ Interaction}
\label{sect:KKstar}

The $K \bar{K}^*$ interaction has been studied by solving the Schr\"odinger equation in the S-wave approximation, and a bound state and a resonance state of $K \bar{K}^*$  are obtained as solutions of the Schr\"odinger equation under different boundary conditions of the wave function, which are identified with the particles $f_1(1285)$ and $f_1(1420)$ in the PDG data, respectively\cite{sunbx2024}.

\begin{figure}
\includegraphics[width=0.6\textwidth]{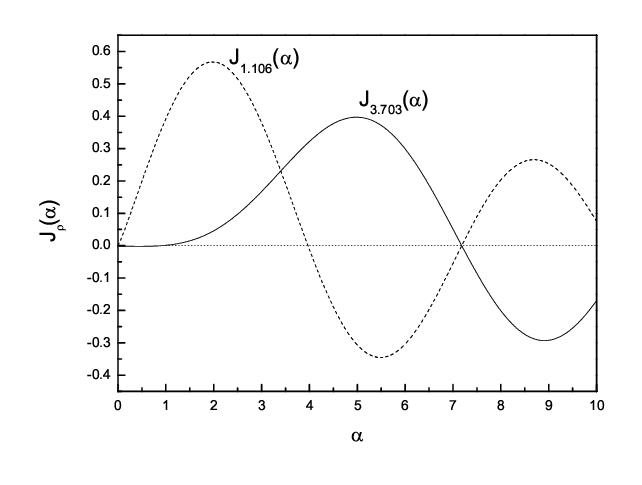}
\caption{
Bessel functions $J_\rho(\alpha)$ with $\rho=3.703$(Solid line) and $\rho=1.106$(Dash line) are depicted, respectively.
}
\label{fig:Bessel}
\end{figure}

The particle $f_1(1285)$ is treated as a deep bound state of $K\bar{K}^*$ with a binding energy of $105$ MeV, so the Bessel function with $\rho=3.703$, which appears as a solution of the Schr\"odinger equation in the Yukawa-type potential, has a zero point at $\alpha=7.1831$.
Its value is larger than the second nonzero zero point of the zeroth order Bessel function $J_0(\alpha)$, which is located at $\alpha=5.520$, so there exists another bound state of $K\bar{K}^*$ in addition to the particle $f_1(1285)$.
Since the Bessel function $J_\rho(\alpha)$ with an order of $\rho=1.106$ is zero at $\alpha=7.1831$, as shown in Fig.~\ref{fig:Bessel}, the binding energy of another $K\bar{K}^*$ bound state can be determined according to Eq.~(\ref{eq:rhoenergy}), which is about $9$ MeV lower than the threshold of $K\bar{K}^*$, and has the same spin and parity as those of the particle $f_1(1285)$. 
This weak bound state near the $K\bar{K}^*$ threshold has not been listed in the PDG manual, and for convenience, it is labeled $f_1(1378)$ in this manuscript. Undoubtedly, it would be the task of experimental scientists to discover it in the future.

The first nonzero zero point of the Bessel function $J_{1.106}(\alpha)$ lies at $\alpha=3.9696$, while $\alpha=7.1831$ is the second nonzero zero point of it. It is apparent that the resonance state of $K\bar{K}^*$ can also be determined according to the outgoing wave condition in Eq.~(\ref{eq:outgoing})
with $\alpha=3.9696$. Along this clue, another resonance state of $K\bar{K}^*$ is produced, which lies at $1425-i41$MeV in the complex energy plane in the center-of-mass frame. Apparently, its mass and width are close to those of the resonance state obtained with the zero point at $\alpha=7.1831$, respectively, as listed in Table.~\ref{table:KKstar}, so it can be assumed that these two resonance states obtained with different zero points correspond to the same particle $f_1(1420)$ in the PDG data. The energies of the $K\bar{K}^*$ resonance states and the properties of the $f_1(1420)$ particle are summarized in Table~\ref{table:KKstar}.

According to Eq.~(\ref{eq:couplingzeropoint}), different zero point values correspond to different coupling constants $g$ in the Yukawa-type potential. The coupling constant $g=1.682$ is obtained with the zero point at $\alpha=7.1831$, while a weaker coupling constant $g=0.9293$ is related to the first non-zero zero point at $\alpha=3.9696$ of the Bessel function $J_{1.106}(\alpha)$. 

\begin{table}[htbp]
 \renewcommand{\arraystretch}{1.2}
\centering
\vspace{0.5cm}
\begin{tabular}{c|c|c|c|c|c|c}
\hline\hline
B(MeV)& $\alpha$ & energy(MeV) &   name & $I^G(J^{PC})$ &  mass(MeV) & width(MeV) \\
 \hline
$105$ &$7.1831$ & $1417-i18$ & $f_1(1420)$ & $0^+(1^{++})$ &   $1428.4^{+1.5}_{-1.3}$  & $56.7\pm3.3$
  \\
$9$ &$3.9696$ & $1425-i41$ &  &  &     & 
  \\  
\hline\hline
\end{tabular}
\caption{
The complex energy of the possible $K \bar{K}^*$ ($\bar{K} K^*$) resonance state with the outgoing wave condition in Eq.~(\ref{eq:outgoing}) and the possible corresponding particle  in the PDG data. The binding energies $B$ of $f_1(1285)$ and $f_1(1378)$ as $K \bar{K}^*$ ($\bar{K} K^*$) bound states are listed, respectively. 
}\label{table:KKstar}
\end{table}

\section{$D \bar{D}^*$ Interaction}
\label{sect:DDstar}

In this section, this method is extended to the study of the $D \bar{D}^*$ system.
If the $D \bar{D}^*$ interaction is realized by exchanging a pion, in order to determine the coupling constant $g$, the particle $\chi_{c1}(3872)$ is assumed to be a bound state of $D^0 \bar{D}^{*0}$ or $\bar{D}^0 D^{*0}$. Since the mass of the particle $\chi_{c1}(3872)$ lies almost at the threshold of $D^0 \bar{D}^{*0}$, the order of the Bessel function is zero according to Eq.~(\ref{eq:rhoenergy}). Therefore, the coupling constant of $D^0 \bar{D}^{*0}$ is related to the first zero point of the Bessel function $J_0(\alpha)$, which is located at $\alpha=2.405$, and thus the coupling constant in the potential of $D^0 \bar{D}^{*0}$ can be obtained according to Eq.~(\ref{eq:couplingzeropoint}), $g=0.323$.
With the zero point at $\alpha=2.405$, the order of the second kind of Hankel function can be obtained by solving Eq.~(\ref{eq:outgoing}), which is complex and relevant to the energy and decay width of the resonance state of $D^0 \bar{D}^{*0}$. The corresponding energies of the resonance states of $D^0 \bar{D}^{*0}$ or $\bar{D}^0 D^{*0}$ are summarized in Table~\ref{Table:DDstar}, and the corresponding particles in the PDG review are also listed.

\begin{figure}
\includegraphics[width=0.5\textwidth]{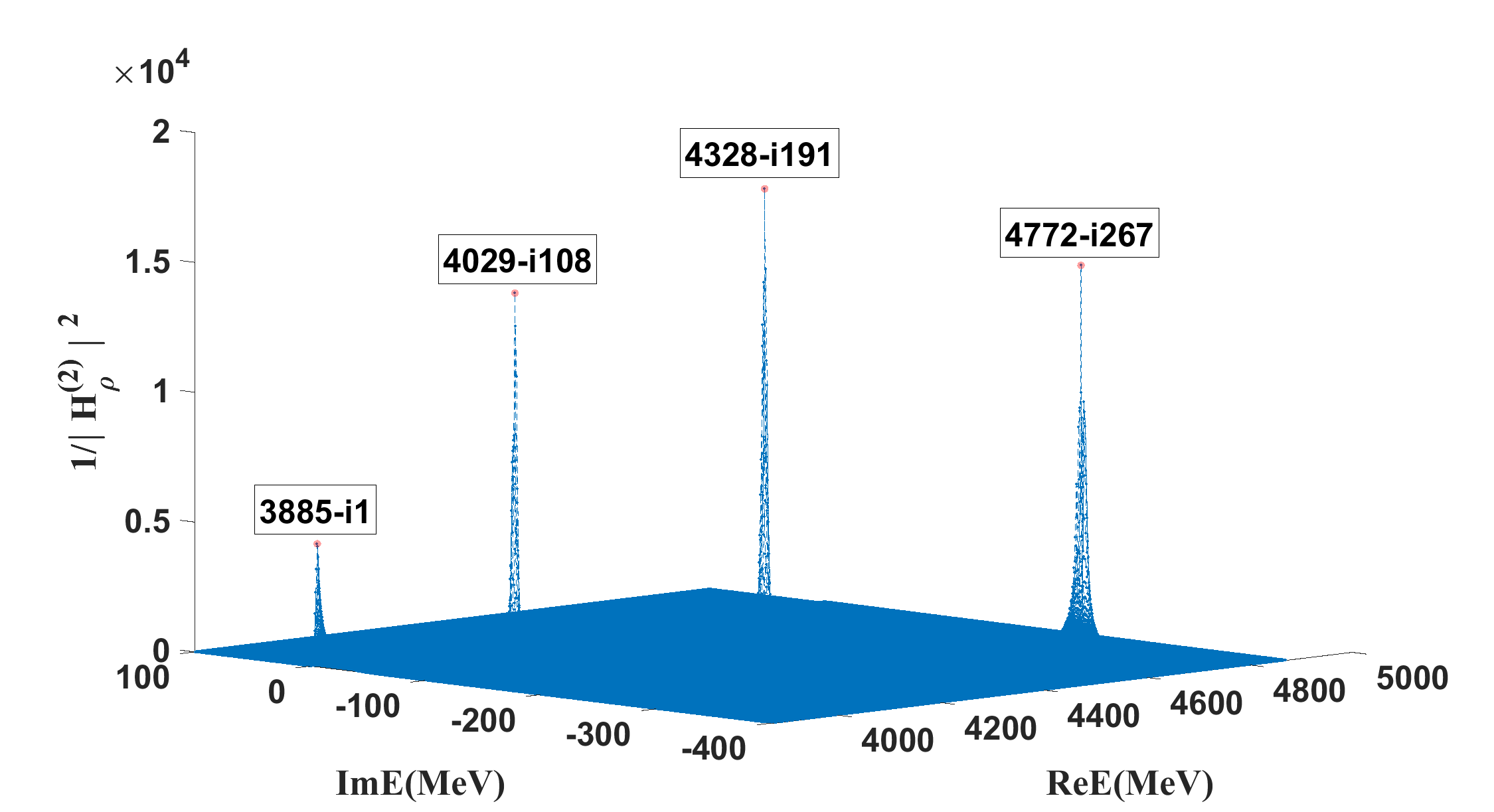}
\includegraphics[width=0.5\textwidth]{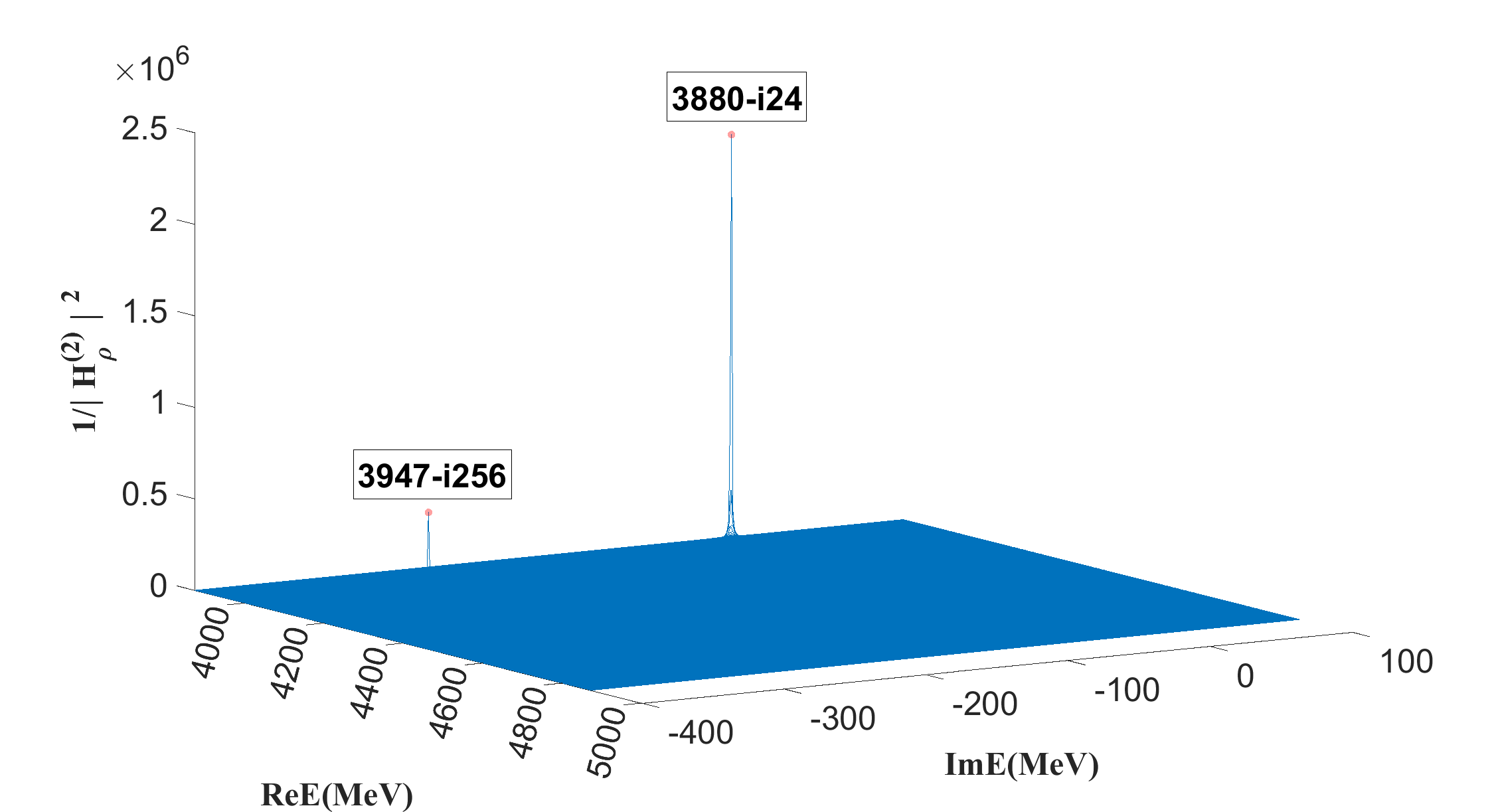}
\caption{$1/|H_\rho^{(2)}(\alpha)|^2$ .vs. the complex energy $E$. $D^0 \bar{D}^{*0}$ channel with the zero point at $\alpha=2.405$ (Left panel) and $D^{-} D^{*+}$ channel with the zero point at $\alpha=4.8583$ (Right panel). The pole of $1/|H_\rho^{(2)}(\alpha)|^2$ corresponds to a zero-point of the second kind of Hankel function $H_\rho^{(2)}(\alpha)$, which represents a $D \bar{D}^{*}$ resonance state, as labeled in the figure.}
\label{fig:ddstar}
\end{figure}

There are four resonance states generated by solving Eq.~(\ref{eq:outgoing}), as depicted in the left panel of Fig.~\ref{fig:ddstar}. The first is located at $3885-i1$ MeV in the complex energy plane, which is above the $D^0 \bar{D}^{*0}$ threshold and could correspond to the particle $T_{c\bar{c}1}(3900)$. 
In addition to the resonance state at $3885-i1$MeV, there are other three resonance states generated dynamically at $4029-i108$MeV, $4328-i191$MeV and $4772-i267$MeV as listed in Table~\ref{Table:DDstar}. 
The mass of the state at $4029-i108$MeV is close to the particle $T_{c \bar{c}}(4020)$, therefore, it is reasonable to assume that the resonance state at $4029-i108$MeV could be associated with the particle $T_{c \bar{c}}(4020)$. If so, the spin and parity of the particle $T_{c \bar{c}}(4020)$ would be determined as $J^P=1^+$.
Moreover, the resonance states at $4328-i191$ MeV and $4772-i267$ MeV 
are considered to correspond to the particles $\chi_{c1}(4274)$ and $\chi_{c1}(4685)$, respectively.

\begin{table}[htbp]
 \renewcommand{\arraystretch}{1.2}
\centering
\vspace{0.5cm}
\begin{tabular}{c|c|c|c|c|c}
\hline\hline
 $D^0 \bar{D}^{*0}$ & energy &   name & $I^G(J^{PC})$ & mass & width \\
 \hline
1 & $3885-i1$ & $T_{c\bar{c}1}(3900)$ & $1^+(1^{+-})$ &  $3887.1\pm2.6$  & $28.4\pm2.6$
  \\
2 & $4029-i108$  & $T_{c\bar{c}}(4020)$ & $1^+(?^{?-})$    & $4024.1\pm1.9$ & $13\pm5$ \\                       
3 & $4328-i191$  &  $\chi_{c1}(4274)$ & $0^+(1^{++})$ &   $4286^{+8}_{-9}$ & $51\pm7$  \\                           
4 & $4772-i267$  & $\chi_{c1}(4685)$ &  $0^+(1^{++})$ &   $4684^{+15}_{-17}$ & $126^{+40}_{-44}$  \\
\hline\hline
\end{tabular}
\caption{
The complex energies of the possible $D^0 \bar{D}^{*0}$ ($\bar{D}^0 D^{*0}$) resonance states with the outgoing wave condition in Eq.~(\ref{eq:outgoing}) and the possible corresponding particles in the PDG data. All are in units of MeV.
}\label{Table:DDstar}
\end{table}

\begin{table}[htbp]
 \renewcommand{\arraystretch}{1.2}
\centering
\vspace{0.5cm}
\begin{tabular}{c|c|c|c|c|c}
\hline\hline
 $D^- D^{*+}$ & energy &   name & $I^G(J^{PC})$ & mass & width \\
 \hline
1 & $3880-i24$ & $T_{c\bar{c}1}(3900)$ & $1^+(1^{+-})$ &  $3887.1\pm2.6$  & $28.4\pm2.6$
  \\
2 & $3947-i256$  & $X(3940)$ & $?^?(?^{??})$    & $3942^{+9}_{-8}$ & $37^{+27}_{-17}$ \\                       
\hline\hline
\end{tabular}
\caption{
The complex energies of the possible $D^- D^{*+}$ ($D^+ D^{*-}$) resonance states with the outgoing wave condition in Eq.~(\ref{eq:outgoing}) and the possible corresponding particles in the PDG data. All are in units of MeV.
}\label{Table:DDstar2}
\end{table}

The particle $\chi_{c1}(3872)$ is 8.11 MeV below the threshold of $D^{-} D^{*+}$ or $D^{+} D^{*-}$\cite{PDG2024}, so the $\chi_{c1}(3872)$ particle can also be considered as a bound state of $D^{-} D^{*+}$ or $D^{+} D^{*-}$ with a binding energy of 8.11 MeV. Therefore, the order of the Bessel function in Eq.~(\ref{eq:Besselzeropoint}) takes a value of $\rho=1.796$, and the first non-zero zero point of the corresponding Bessel function lies at $\alpha=4.8583$. According to Eq.~(\ref{eq:couplingzeropoint}), the coupling constant of $D^{-} D^{*+}$ in the Yukawa-type potential can be determined, which takes a value of $g=0.6520$ and is approximately twice the coupling constant of $D^0 \bar{D}^{*0}$.
With the zero point at $\alpha=4.8583$, the energy and decay width of the resonance state can be obtained when the outgoing wave condition in Eq.~(\ref{eq:outgoing}) is taken into account. Consequently, two resonance states are generated as solutions of the Schr\"odinger equation, which are located at $3880-i24$MeV and $3947-i256$MeV in the complex energy plane, respectively, as shown in the right panel of Fig.~\ref{fig:ddstar}.
It is reasonable to assume that the resonance state at $3880-i24$MeV corresponds to the particle $T_{c\bar{c}1}(3900)$, while the resonance state at
$3947-i256$MeV represents the particle $X(3940)$ in the PDG data\cite{PDG2024}. In the S-wave approximation, the spin and parity of the particle $X(3940)$ can be determined as $J^P=1^+$. The resonance states of $D^{-} D^{*+}$ or $D^{+} D^{*-}$ and the corresponding particles in the PDG review are listed in Table~\ref{Table:DDstar2}.

\section{Summary}
\label{sect:summary}

The systems of $K \bar{ K}^*$ and $D \bar{ D}^*$ are studied by solving the Schr\"odinger equation with different boundary conditions. The one-pion-exchange potential is assumed to play an important role in the interaction of these two systems. By fitting the coupling constant with the binding energy of the bound state, some resonance states are obtained as solutions of the Schr\"odinger equation when the outgoing wave condition is taken into account, and most of them are interpreted as corresponding to the particles in the PDG data, respectively. Therefore, it is concluded that there are intrinsic relations between the bound state and the resonance state of the system.


\begin{thebibliography}{99}

\bibitem{PDG2024}
S.~Navas \textit{et al.} [Particle Data Group],
Phys. Rev. D \textbf{110}, 030001 (2024)



\bibitem{Roca2005}
L.~Roca, E.~Oset and J.~Singh,
Phys.\ Rev.\ D {\bf 72}, 014002 (2005)

\bibitem{sun1420}
D.~M.~Wan, S.~Y.~Zhao and B.~X.~Sun,
arXiv:1808.08358 [hep-ph].


\bibitem{sunbx2024}
B.~X.~Sun, Q.~Q.~Cao and Y.~T.~Sun,
Commun. Theor. Phys. \textbf{76}, 105301 (2024)


\bibitem{Moiseyev}
N. Moiseyev, {\it Non-Hermitian Quantum Mechanics}, Cambridge University Press, New York, 2011.

\end{thebibliography}
\end{document}